\documentclass[aps,pre,floatfix,twocolumn,footinbib,showpacs]{revtex4-1} 
\usepackage{epsfig} 
\usepackage{mathtools}
\usepackage{comment}
\usepackage{hyperref}
\begin{document}
\title{Vertical probe-induced asymmetric dust oscillation in complex plasma}
\author{B.~J.~Harris}
\email{Electronic mail: brandon\_harris@baylor.edu}
\author{L.~S.~Matthews}
\email{lorin\_matthews@baylor.edu}
\author{T.~W.~Hyde}
\email{truell\_hyde@baylor.edu}
\affiliation{CASPER (Center for Astrophysics, Space Physics, and Engineering Research)\\Baylor University, Waco, Texas 76798-7310, USA}
\date{\today}
\begin{abstract}

Spherical, $\mu$m-sized particles within a Coulomb crystal levitated in the sheath above the powered lower electrode in a GEC reference cell are perturbed using a Zyvex S100 Nanomanipulator. Using the S100, a vertical probe is positioned within the cell at various locations with respect to the crystal formed within the sheath. As the probe is lowered toward the horizontal plane of the dust layer, a circular cavity opens in the center of the crystal and expands. To explore the minimally perturbative state, the probe is lifted to the position that closes this cavity, the probe potential is oscillated, and the motion of the particle directly beneath the probe is analyzed. Using a simple electric field model for the plasma sheath, the change predicted in the levitation height is compared with experiment.

\end{abstract}


\maketitle
\section{\label{sec:Int}Introduction}

When macroscopic dust is introduced into a plasma, the resulting complex plasma system offers the opportunity to simulate, at its most fundamental level, interesting physics pertaining to a host of research areas, including protoplanetary \cite{ilgner} or protostellar \cite{jones} development, contamination in plasma enhanced semiconductor atomic layer deposition and etching systems \cite{vitale}, and wall erosion within a fusion device \cite{liu}. At low temperature and power, it also provides a basis for the study of entirely new plasma physics.

A recurring problem in complex plasma physics is obtaining accurate measurements in the sheath region, which is nonuniform and spatially limited compared to the bulk. Dust particles in a capacitively coupled rf plasma levitating beneath the bulk plasma can act as \emph{in situ} probes providing insight into the fundamental sheath parameters \cite{swinkels}. One difficulty with this method is accurately determining the charge on the dust particles, which is itself linked to the plasma parameters. Given known plasma conditions, Orbital Motion Limited (OML) theory \cite{allen} can be used to predict the amount of charge collected by each particle. An improved charge estimation can be obtained using a shifted Maxwellian distribution to incorporate the effect of ions streaming toward the lower electrode \cite{khrapak}, although this requires an estimation of the ion velocity which is also not easily measured.

A plate with a milled depression 1-mm deep placed on the lower electrode can provide horizontal confinement, allowing dust particles to be stably levitated in a GEC rf cell through a balance of the vertical forces. In this case, the dust can achieve a minimum energy state by organizing into a two-dimensional circular lattice with hexagonal symmetry.

Many methods have been used to manipulate the dust in such a complex plasma crystal. These include modifying the dc bias on the lower electrode \cite{trottenberg}, applying radiation pressure \cite{homann2}, initiating a wave through the crystal using a powered horizontal wire \cite{samsonov}, instituting a temperature gradient to establish a thermophoretic force \cite{rothermel}, altering the particle charge or sheath electric field by changing basic plasma parameters such as the power and/or neutral gas pressure \cite{fortov}, and beaming a pulsed stream of ions \cite{wiese}.

\begin{figure}
\includegraphics[scale=0.63,trim=2 0 0 0]{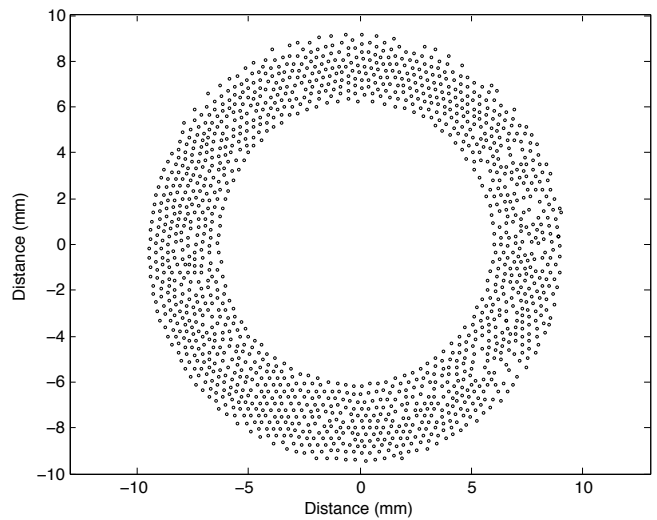}
\caption{Dust particle positions in a planar dust crystal with an open central cavity created by a negatively biased vertical probe located at the center of the cavity region. This was achieved using a system power of 10 W, a neutral gas pressure of 100 mTorr, a probe potential of -30 V, and a 1-inch diameter circular cutout for horizontal confinement.
\label{void}} 
\end{figure}

The addition of a probe to the system provides the ability to apply a controlled potential difference at various positions relative to the charged dust layers. Previous experiments have used such probes to induce voids in a DC plasma \cite{thomas} and in microgravity rf plasmas \cite{klindworth}. 

In this experiment, an adjustable probe is positioned perpendicularly to a horizontal dust lattice. As it approaches the crystal plane, the particles move radially outward, creating an open circular space within the crystal as shown in Fig.~\ref{void}. This will henceforth be referred to as the cavity. Under operating conditions that just close the cavity, a numerical model is used to calculate the charge on the dust particle. The probe potential is oscillated, and the resultant dust motion is analyzed to determine the neutral drag coefficient and the resonant frequency. The plasma parameters within the model are then varied to find the best fit to observed particle oscillations. Assuming probe interaction with the dust is a superposition of near (i.e., direct interaction with the probe potential) and far (i.e., changes to the plasma) field effects, this experiment isolates the latter.

\begin{figure}
\includegraphics[scale=1]{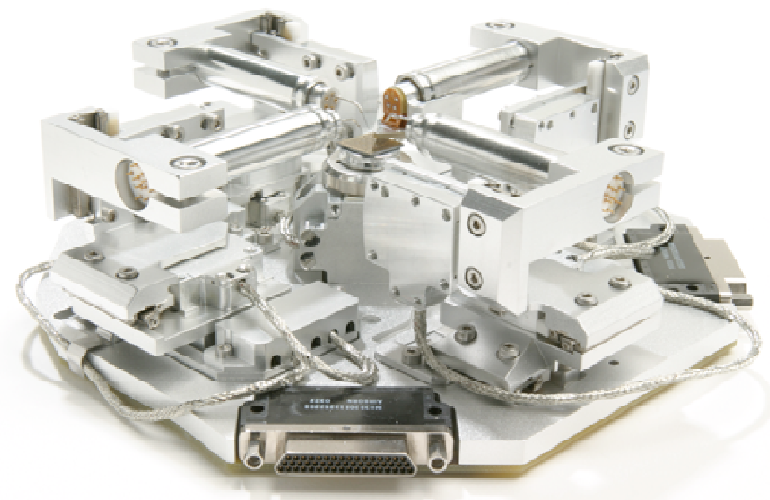}
\caption{(Color online) The Zyvex S100 Nanomanipulator head is located inside the plasma cell, mounted upside down above the upper ring electrode. One of the four manipulators shown is used in this experiment to position the probe as described in the text.\label{zyvexnano}} 
\end{figure}

\begin{table}
\begin{tabular}{|c|c|}
\hline
Parameter & Settings\\
\hline
Probe Bias & 10, \textbf{20}, 30 V\\
Probe Peak to Peak & 45, \textbf{55}, 65 V\\
System Power & 0.75, \textbf{1.00}, 1.50 W\\
Pressure & 70, \textbf{80}, 90 mTorr\\
Probe Height & \textbf{7}, 9, 11 mm\\
Frequency & 1, \textbf{2.3}, 5 Hz\\
DC Bias & -10, \textbf{-5}, -1 V\\
\hline
\end{tabular}
\caption{Experimental parameters used in this experiment. The bolded value for each parameter defines the base configuration, as discussed in the text.}
\label{parameters}
\end{table}

The arrangement of this paper is as follows: First, the experimental apparatus is described. Second, the experiment itself is introduced and forced particle oscillation over a range of parameters is explored. Third, a simple electric field model is applied to predict the equilibrium levitation heights of the crystals. Different parameters of this model are then perturbed to incorporate the oscillation, and the results are compared to experiment. Finally, the conclusions are reported.

\section{\label{sec:App} Apparatus}

A modified GEC (Gaseous Electronics Conference) rf reference cell \cite{hargis, land} located at the Center for Astrophysics, Space Physics, and Engineering Research (CASPER) was employed for this study. The cell contains two electrodes, 8 cm in diameter, separated by 1.9 cm. The lower electrode is powered at 13.56 MHz while the upper ring-shaped electrode and chamber are grounded. In this experiment 8.9 micron diameter melamine formaldehyde (MF) dust particles were levitated in an argon plasma. A plate containing a circular cutout 1 mm deep and 25.4 mm (1 inch) in diameter was placed on the lower electrode to provide horizontal confinement of the dust particles. Images of the dust particles were captured at 125 and 250 frames per second using a side-mounted CCD camera. Table \ref{parameters} lists the system parameters used for this experiment. The actual power delivered to the plasma, calculated by measuring the phase difference between the current and the voltage, was 0.23, 0.36, and 0.56 W ($V_{rf}$=9.6, 11.4, and 14.1 V). The DC bias of the lower electrode was held fixed using an external power supply. Langmuir probe measurements of the bulk plasma were also collected using a SmartProbe (Scientific Systems Ltd) \cite{land}.

\begin{figure}
\includegraphics[scale=0.425]{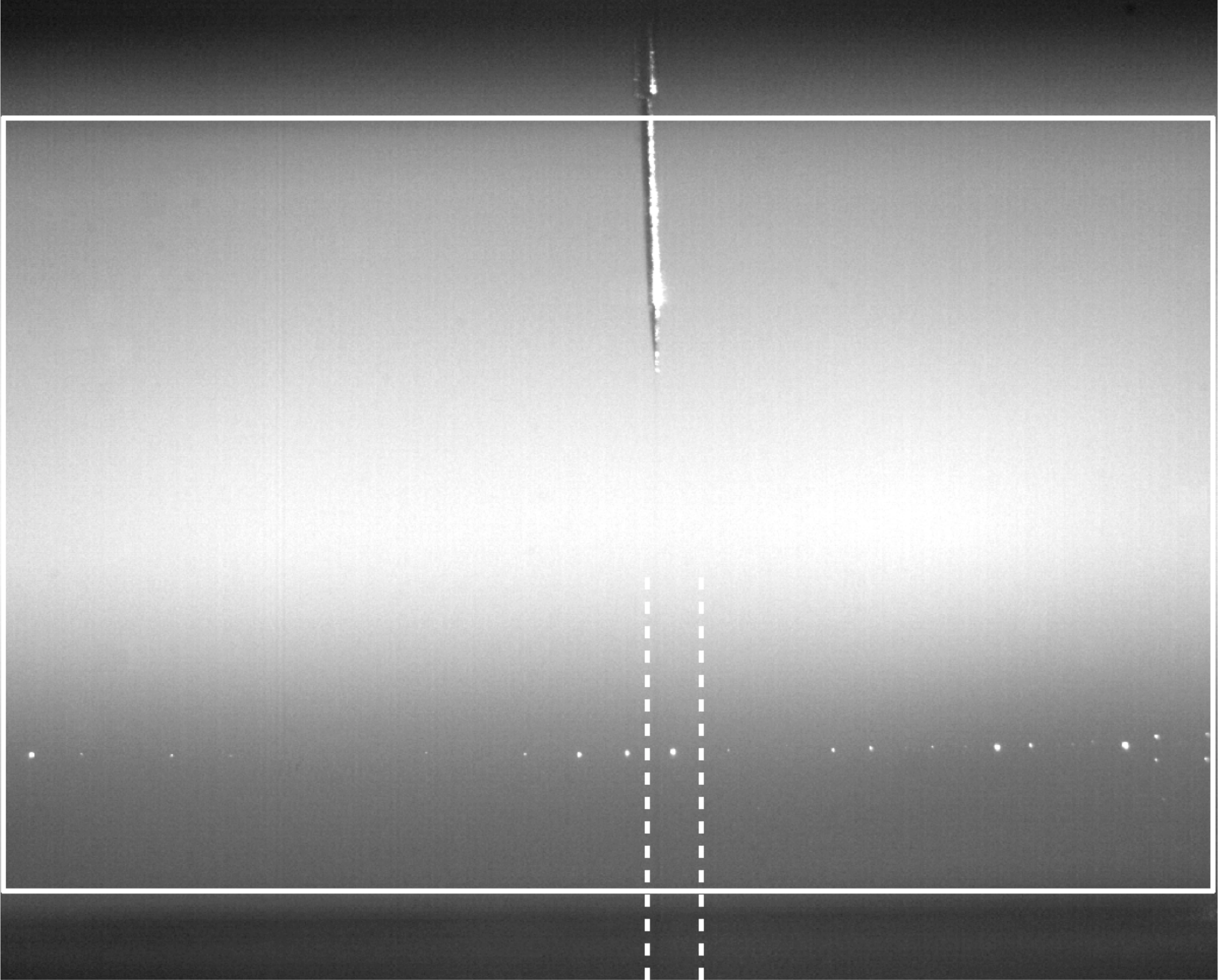}
\caption{Side view of the vertical oscillation experiment. For the `base configuration' setup, the tip of the probe is located 7.3 mm above the dust equilibrium levitation height. Only the particle between the (dashed) bars is tracked, with the area in the box, which excludes the electrodes, used for calibration.\label{sideview}}
\end{figure}

The Zyvex S100 \cite{zyvex} head (originally designed for sample manipulation in microscopy applications) was attached within the plasma chamber to act as a perturbation tool (Fig.~\ref{zyvexnano}). A hollow cylindrical probe 48 mm in length, having an outer diameter of 450 $\mu$m, was connected to the S100 head. A tip, with an adjustable length of up to 14 mm and a diameter tapering from 250 to 50 microns over the last 100 microns, protruded from the probe (Fig.~\ref{sideview}). The S100 is capable of remote controlled movement of the probe by up to 10 mm in all three dimensions. The probe potential can be controlled with respect to the ground using an external power supply.

\section{Experiment\label{sec:Exp}}

Motivation for the current experiment stemmed from the observation that changing the probe bias and height both produced and altered the size of a circular cavity in the dust layer. In the horizontal wire experiment referenced above \cite{samsonov}, the primary repulsion force upon the dust was shown to be produced by the direct electric field. This proves not to be the case here. In all cases the cavity created is much larger than would be expected assuming a simple interaction between the probe and the dust through a shielded Coulomb potential. In order to examine this behavior, the probe tip was positioned at the transition height, defined as the point of closest approach before a cavity is opened in the crystal. A side view showing the probe at the transition height for a crystal at its equilibrium levitation height for the base configuration as defined in Table \ref{parameters} is shown in Fig.~\ref{sideview}.

\begin{figure}
\includegraphics[scale=0.585,trim=5 0 5 0]{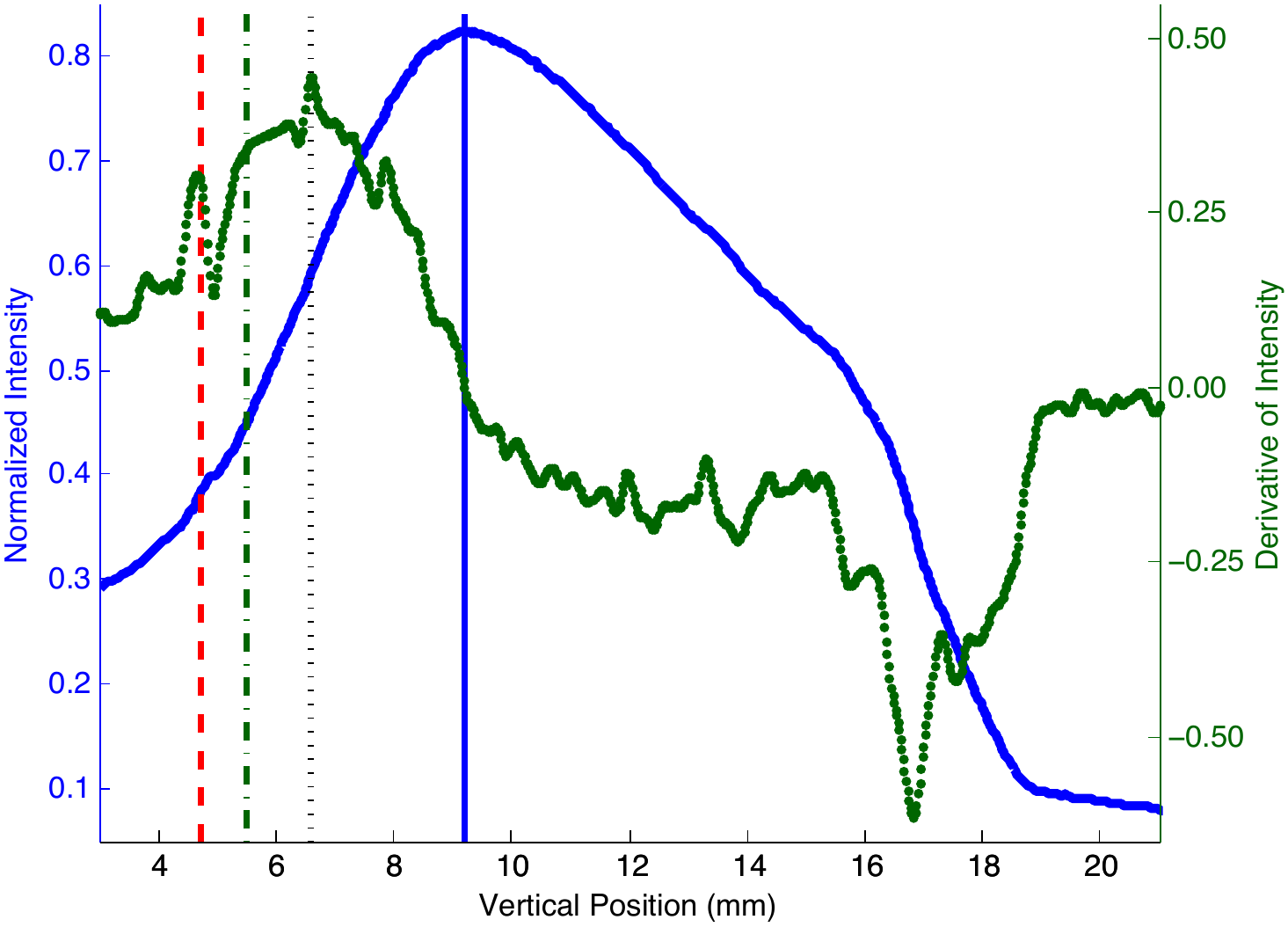}
\caption{(Color online) Normalized optical emission intensity profile (solid blue line) of the plasma (Fig.~\ref{sideview}) averaged over the horizontal coordinate as a function of the distance above the lower electrode (where the upper electrode is located at $z$=19 mm). The derivative of the intensity (green dots) is superimposed with vertical lines indicating the position of the dust (red dashed), the 1/$e$ point (green dash-dot), the local maximum of the derivative emission intensity (dotted black), and the maximum of the intensity (solid blue).\label{intprof}}
\end{figure}

\begin{figure}
$\begin{array}{ccc}
\includegraphics[scale=0.757,trim=3 0 0 0]{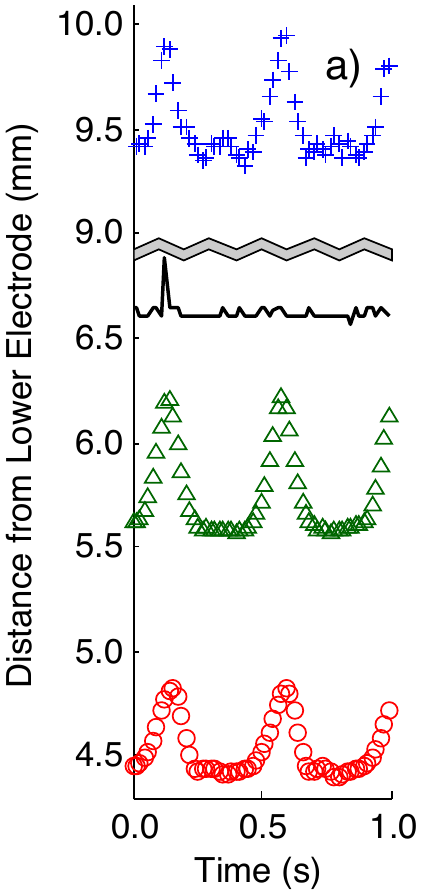} &
\includegraphics[scale=0.757,trim=14 0 0 0]{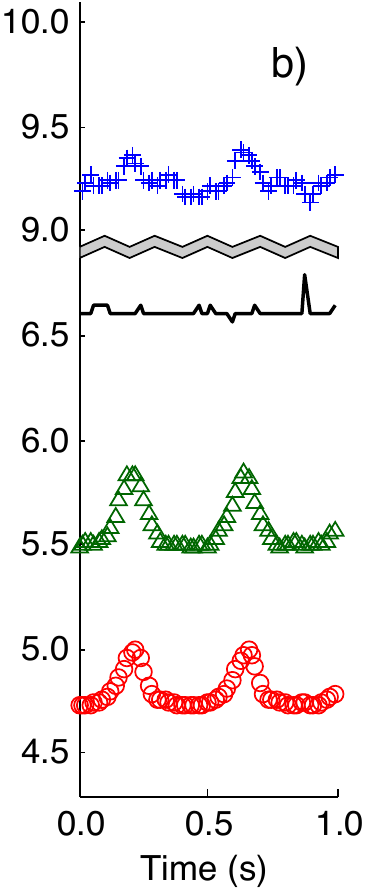} &
\includegraphics[scale=0.757,trim=14 0 0 0]{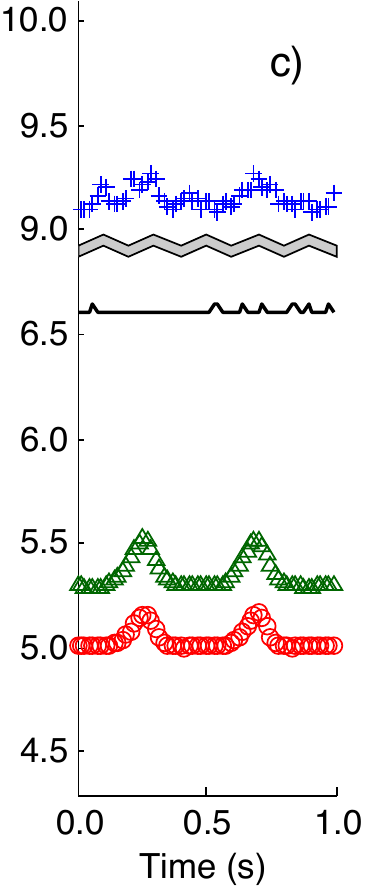} 
\end{array}$
\caption{(Color online) Distance from the lower electrode versus time for: the particle height (red circles), the sheath edge (green triangles), the maximum of the derivative of the emission intensity (black line), and the maximum of the emission intensity (blue crosses), for system powers of a) 0.75 W, b) 1.00 W, and c) 1.50 W.\label{vertoscpos}}
\end{figure}

Datasets consisting of 125 images taken at 125 fps were obtained for each of the parameters shown in Table \ref{parameters}. The set of parameters in bold (20 V probe bias, 55 V probe potential peak to peak, 1.0 W system power, 80 mTorr gas pressure, 7.3 mm probe height, 2.3 Hz oscillation frequency, and -5 V fixed DC bias) will henceforth be referred to as the base configuration. As each parameter is modified in the experiment, all others remain the same as those established for the base configuration.

Images were analyzed employing both particle tracking \cite{partracker} and profile analysis \cite{imagej}. Image calibration was completed by subtracting the minimum intensity of the image set in the region excluding the electrodes (indicated by the box in Fig.~\ref{sideview}) from all pixels before determining the position of the maximum intensity and sheath edge. The probe was removed from the image before profile analysis by copying the adjacent region.

The quantities best characterizing the oscillation are indicated in a representative image profile of the base configuration in Fig.~\ref{intprof}. These include, in ascending vertical position, the particle levitation height, the sheath edge (defined as the position where the optical emission intensity decreases by a factor of $e$ from its maximum \cite{beckers}), the location of the maximum derivative of the emission intensity in the lower sheath, and the point of emission maximum. Sinusoidal oscillation of the probe potential was found to yield non-sinusoidal change in these four values, as shown in Fig.~\ref{vertoscpos}, over two oscillation cycles for three different system powers.

The fact that the maximum emission intensity moves lower with an increase in power is most likely due to cell geometry; the plasma exhibits an asymmetric vertical emission profile (averaged over the horizontal direction) as shown in Fig.~\ref{intprof}, due in part to the fact that the upper electrode is a hollow ring and grounded whereas the lower electrode is a powered plate. Oscillation of the probe potential causes the change in amplitude of the emission intensity maximum to become significant at lower powers (\ref{vertoscpos}a), but reduced at higher powers (\ref{vertoscpos}c).

It was found that the position of the local maximum of the derivative of the emission profile of the lower sheath remains constant with change in power (Fig.~\ref{vertoscpos}). This is true not only for variations in power, but over all parameters tested. As shown in Fig.~\ref{intprof}, the local maximum is located at $z$=6.6 mm, with a local minimum at $z$=16.8 mm in the upper sheath. The peak which can be seen at $z$=4.7 mm is due to the line of dust.

The location of the 1/$e$ point, indicating the sheath edge, decreases with an increase in power while the particle levitation height increases (Fig.~\ref{vertoscpos}), both due to an increase in ionization of the plasma. The definition of sheath edge position used in this experiment was first proposed by Beckers et al.~\cite{beckers}. An alternative experimental method to determine the sheath edge based on theory is measurement of the equilibrium height of nanoparticles \cite{samarian}. Applying the definition of the sheath edge as the 1/$e$ point to Fig.~4 in \cite{samarian} results in 0.7\% agreement in vertical position, justifying the choice in the present work.

\begin{figure}
\includegraphics[scale=0.643,trim=5 0 0 0]{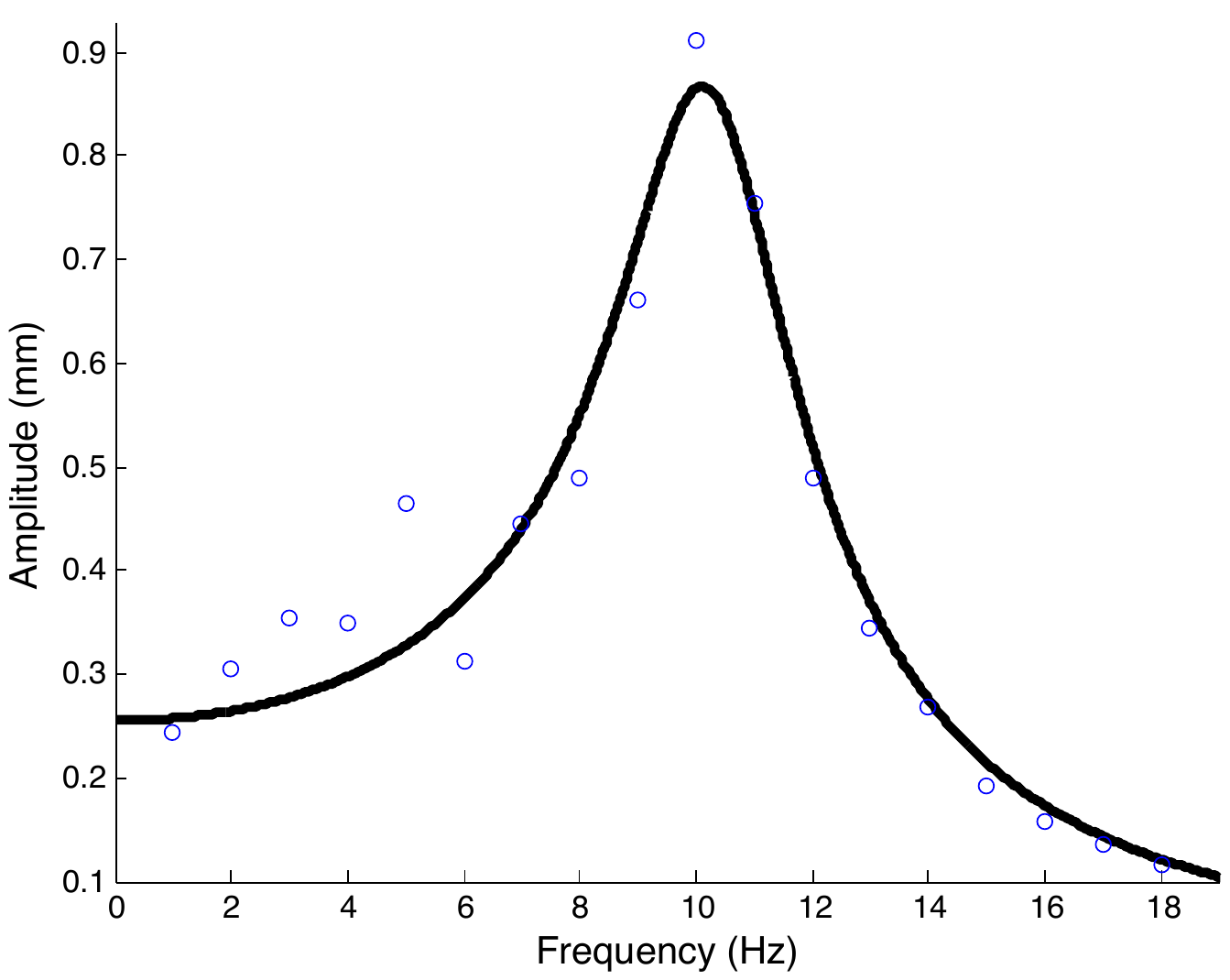}
\caption{(Color online) Particle oscillation amplitude versus probe oscillation frequency. The fit shown is for a damped, forced harmonic oscillator.\label{ampvsfreq}}
\end{figure}

\begin{figure}
\includegraphics[scale=0.63,trim=5 0 0 0]{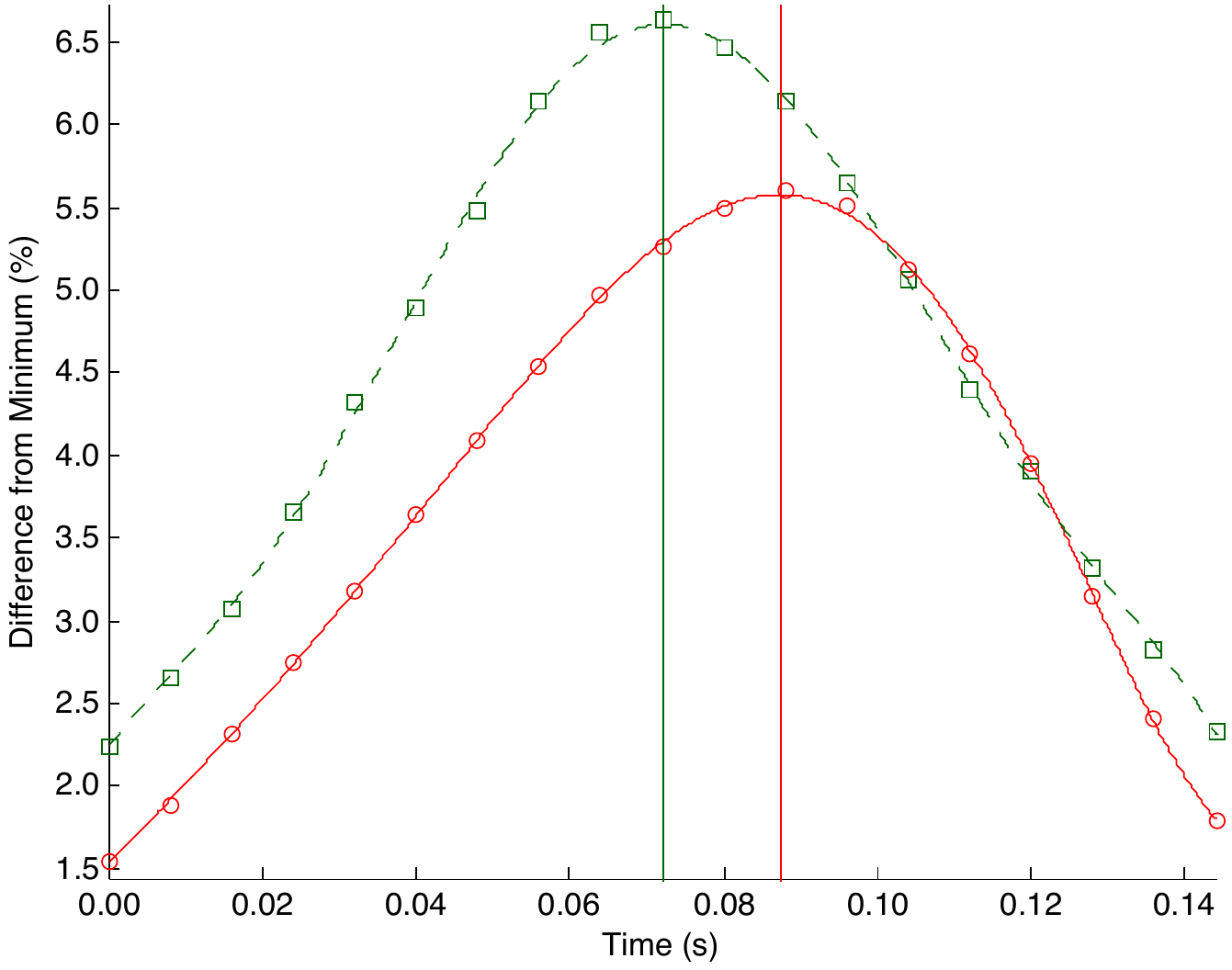}
\caption{(Color online) Expanded view of the base configuration (Fig.~\ref{vertoscpos}b). The solid (red) line is a polynomial fit to the particle position (red circles) over time while the dashed line represents the fit to the sheath position (green squares). The vertical lines shown indicate the peaks of these fit lines illustrating the delay between maximum particle response and maximum sheath response. For this case, the delay is 15.0 ms; delays of similar magnitude were found for other experimental parameters. All values are shown as percent differences from their respective minima.\label{delay}}
\end{figure} 

The response of the dust to the probe oscillation can be approximated as in \cite{trottenberg} by the steady-state response of a damped, forced harmonic oscillator, with the amplitude given by
\begin{equation}
A(\omega)=\frac{F}{[(\omega_0^2-\omega^2)^2+4\beta^2\omega^2]^{1/2}}
\label{ampharmonicosc}
\end{equation}
where $\beta$ is the damping coefficient, $\omega_0$ is the resonant frequency, $\omega$ is the driving frequency, and $F$ is the magnitude of the driving force divided by the mass of the dust particle. A plot of the dust particle amplitude versus frequency (Fig.~\ref{ampvsfreq}) shows good agreement with theory. At a pressure of 80 mTorr, the resulting fit parameters are $\beta$=9.66 $s^{-1}$, $\omega_0$=65.0 rad/s (10.3 Hz), and $F$=1.08 N/kg. These are comparable to the results reported in \cite{zhang}, where $\beta$=8 $s^{-1}$, and $\omega_0$/$2\pi$=13.2 Hz, measured at 66 mTorr. The process that creates the non-sinusoidal particle response in time must not be sensitive to changes to frequencies in this range, and thus the only impact to Eqn.~\ref{ampharmonicosc} occurs in $F$. 

The phase delay which can be seen between the maxima of the sheath edge and the particle positions as shown in Fig.~\ref{delay} is well known for damped harmonic oscillation. For all cases, it is found to be in the correct direction with the particle response lagging the driving force. While the probe potential oscillates sinusoidally, the location of the sheath edge does not, as seen in Fig.~\ref{vertoscpos}. This lends a hypothesis that the force on the particle is not sinusoidal. Though a periodic, non-sinusoidal driving force will yield the same resonant frequency as that derived using Eqn.~\ref{ampharmonicosc} \cite{fortov}, calculation of the phase shifts under different driving forces requires a numerical approach. The phase difference ($\delta$) for a sinusoidal force,
\begin{equation}
\delta=tan^{-1} \left( \frac{2\omega\beta}{\omega_0^2-\omega^2} \right),
\end{equation}
applied to this experiment does not lead to a damping constant (12.7 $s^{-1}$) or resonant frequency (7.7 Hz) consistent with that found from the frequency sweep (Fig.~\ref{ampvsfreq}). Deviation of the experimental data from the fit line at half the resonant frequency shown in Fig.~\ref{ampvsfreq} was also seen in two other experiments. Ivlev et al.~\cite{ivlev} drove vertical dust oscillations, using a powered wire placed below the particle layer and oriented horizontally (i.e., parallel) to the lower electrode. While the amplitude fit deviated from a damped harmonic oscillator above driving potentials of 50 mV peak to peak, notable superharmonic response appeared at driving potentials (4 V peak to peak) substantially higher than those used in this experiment. Homann et al.~\cite{homann} examined particle response to lower electrode DC bias oscillations of square and sine waves. A laser was also pulsed on the particle without bias changes which created a square wave force without sheath modification. Superharmonic response was found only when the force delivered to the dust was not sinusoidal. Therefore the phase shift discrepancy and superharmonic peak are two results that confirm that the force on the particle is not sinusoidal and indicate that changes in the plasma are driving the particle's motion, rather than the sinusoidal potential on the probe.

\begin{figure}
\includegraphics[scale=0.631,trim=1 0 0 0]{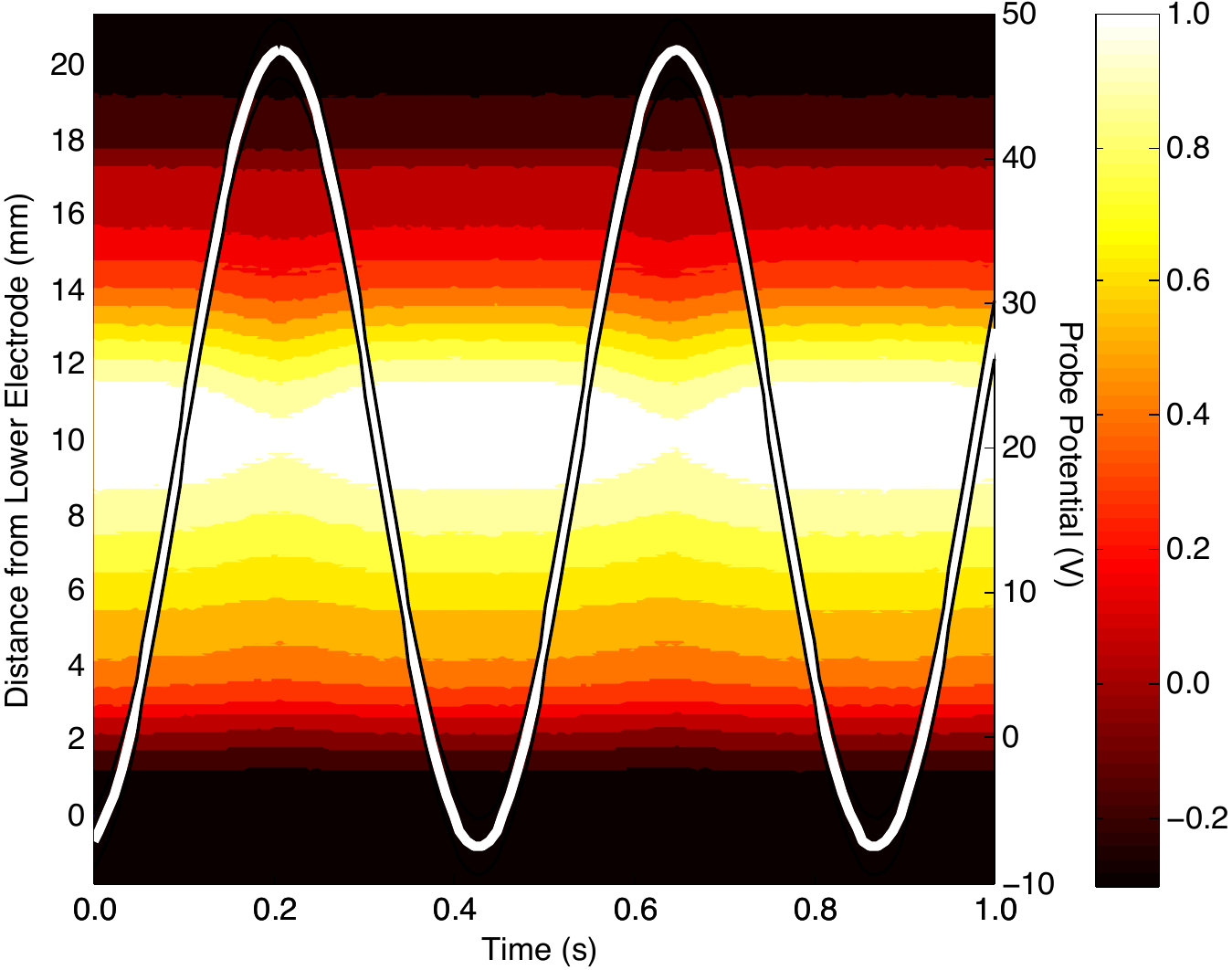}
\caption{(Color online) Intensity space-time contour plot for the base configuration. The colorbar gives the intensity relative to the overall minimum, and normalized to the resulting maximum ($(I-I_{min})/I_{max}$). The sinusoidal probe potential is superimposed.\label{intcontour}}
\end{figure}

\begin{figure}
\includegraphics[scale=0.575,trim=1 0 0 0]{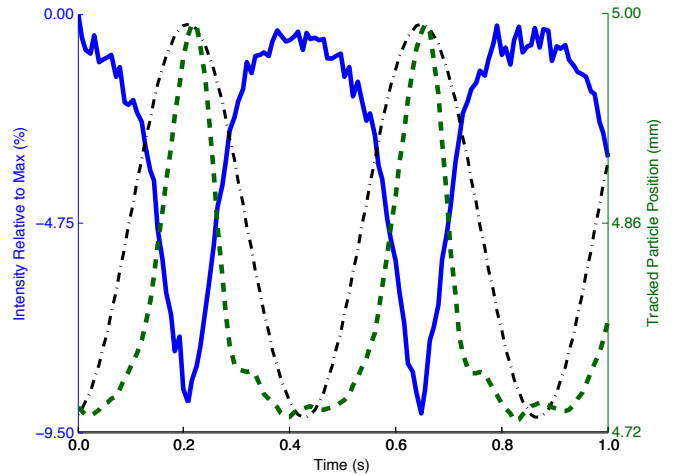}
\caption{(Color online) Maximum intensity percentage change over time for the base configuration (solid blue line), superimposed with tracked particle data (dashed green line) and probe potential (dash-dot black line), which is scaled to the maximum particle position.\label{intparticle}}
\end{figure}

Changes in the plasma can be characterized through the plasma intensity. Fig.~\ref{intcontour} shows the evolution of the plasma intensity over time. The vertical position where the change in intensity is most pronounced is in the plasma bulk. A decrease in the plasma glow results from a reduction in electron density, which in turn results in a reduction of the total argon electron transitions. When the probe reaches its maximum positive potential, the bulk plasma intensity decreases by up to 9\%, as shown in Fig.~\ref{intparticle}. The glow responds faster than the particle, providing an indicator for the force on the particle. This presents a link to the particle delay (Fig.~\ref{delay}) without having to independently match the probe potential to the oscillation. Fixed positive probe potentials corresponded to the dust crystal height being raised as a whole, and the height of the maximum optical emission intensity from the plasma shifted upward as well.

\begin{figure}
\includegraphics[scale=0.645,trim=1 0 0 0]{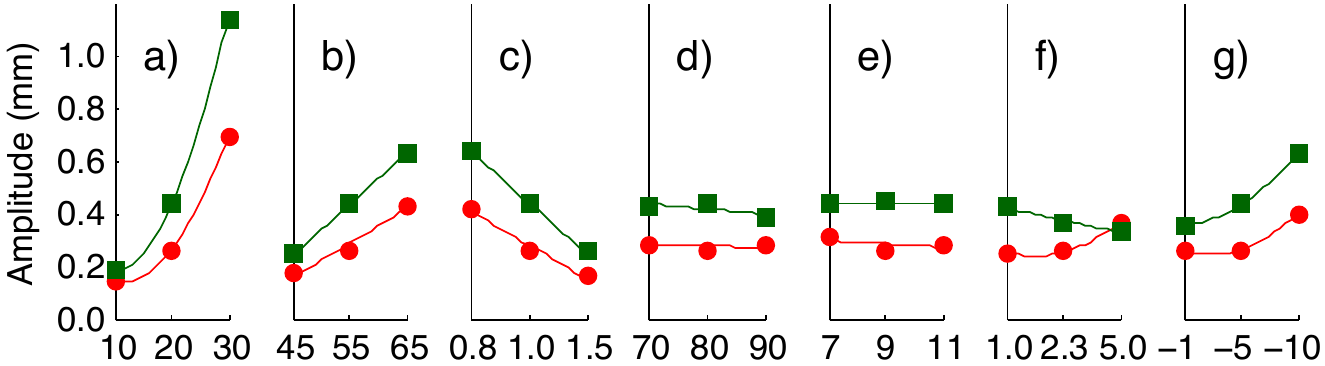}
\caption{(Color online) Oscillation amplitudes for the sheath edge (green squares) and the particle (red circles) with quadratic or linear fits for variation in individual parameters given in Table \ref{parameters}. a) Probe bias (V), b) probe peak to peak (V), c) system power (W), d) pressure (mTorr), e) probe height (mm), f) frequency (Hz), g) DC bias (V).\label{oscamps}}
\end{figure}

The dust oscillation amplitudes for the parameters tested is shown in Fig.~\ref{oscamps}. The amplitude of the sheath oscillation shown is larger than that observed for the particle oscillation. This implies that the particle is not simply entrained in the plasma, but instead that its reduced amplitude is due to its mass (resisting sudden change through inertia); cross-sectional area (motion impeded by colliding neutral gas atoms and ions); charge (Coulomb collisions with the streaming ions); and the structure of the sheath electric field. While any increase in probe bias generates a nonlinear increase in oscillation amplitude (\ref{oscamps}a), an increase in the peak to peak oscillation voltage results in a nearly linear increase (\ref{oscamps}b). On the other hand, increasing the rf power causes a decrease in the oscillation amplitude (\ref{oscamps}c) due to ionization increase and electron temperature decrease (as found by Langmuir probe measurements), both of which increase the overall shielding by decreasing the Debye length. Therefore, a reduced perturbation is applied to the plasma, resulting in a smaller particle amplitude. At the same time, changing the pressure and probe height do not appreciably affect the oscillation amplitude (\ref{oscamps}d and \ref{oscamps}e). As the probe oscillation frequency approaches 5 Hz, the particle no longer has time for relaxation between cycles and the maximum particle amplitude increases due to superposition (\ref{oscamps}f). Decreasing the DC bias on the lower electrode (\ref{oscamps}g) increases the potential difference between the plasma and the electrode, raises the particle within the sheath region, and results in a nonlinear increase in the particle amplitude. Taken together, these results indicate once again that the probe locally modifies the plasma which in turn drives the oscillation of the particle, as supported by the phase delay shown in Fig.~\ref{delay}.

\section{\label{sec:Mod} Model}

A numerical model was developed to examine the data described above. The general equation of motion for a dust particle is given by
\begin{equation}
m\ddot{z}=F_E-F_g-F_i+F_\beta
\end{equation}
where $F_n$ are the forces produced by the electric field, gravity, streaming ions, and neutral damping (whose sign changes so that it always opposes the particle motion), respectively. The electric potential within the sheath is assumed to be parabolic \cite{tomme}, where the fixed DC potential on the lower electrode is $V_0$ at $z$=0 and equal to the plasma potential ($V_p$) at the sheath edge ($z$=d). The electric force acting on a dust particle is then:
\begin{equation}
F_E(z)=\frac{2Q(V_p-V_0)}{d} \left( \frac{z}{d}-1 \right)
\label{eforce}
\end{equation} The charge (Q) on each particle is initially assumed to be 14,750e, where e is the elementary electron charge, as determined by previous experiment in this GEC cell \cite{zhang}. This model was validated by first using it to predict the dust equilibrium levitation height, and then to estimate the charge on a dust particle for each configuration. After the charge values were found, changes in the plasma based on changes of the experimental parameters were explored.

\begin{figure}
\includegraphics[trim=3 0 0 0, scale=0.622]{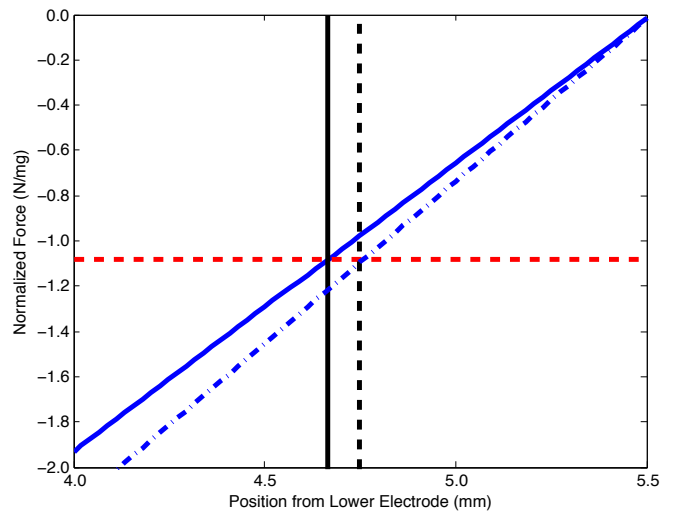}
\caption{(Color online) Forces involved in the levitation of the dust within the plasma sheath. The dashed (red) horizontal line represents the downward force of gravity including ion drag (hence the small deviation from -1). Using the initial estimate for the particle change in Eqn.~\ref{eforce} gives the linear electric field force in the sheath (blue solid diagonal line). The intersection of these two lines gives the predicted equilibrium height (black solid vertical line). The experimentally measured levitation height (black dashed line) can be used to determine the actual particle charge by adjusting the electric force (blue dash-dot line) so that it intersects the combined downward force at the levitation height.\label{forcebalplot}}
\end{figure}

It is well known that ions stream from the plasma bulk and are accelerated into the sheath \cite{chen}. For this experiment, it is assumed these ions exit the sheath edge at the Bohm velocity into a collisionless sheath. However, the ion-neutral mean free path (625 microns at 80 mTorr using the total cross-section from \cite{mcdaniel}) is smaller than the sheath thickness, which means the sheath exhibits some collisionality. Ignoring this effect overestimates the ion drag on a dust particle, since some momentum is transferred to the neutrals; however, this is still quite small relative to the other forces considered \cite{beckers}. Ion drag was calculated using the model of Khrapak et al.~\cite{khrapak} and added to the downward gravitational force. The ion drag calculation was modified to determine the ion and electron densities at the dust position given the electric potential at the dust height, using the densities in the plasma bulk. These modifications included energy conservation and the continuity equation for the ions \cite{lieberman} (which quantifies the decrease in ion density as the ions accelerate toward the lower electrode), and a Boltzmann distribution of electrons.

\begin{figure}
\includegraphics[scale=0.555,trim=2 0 0 0]{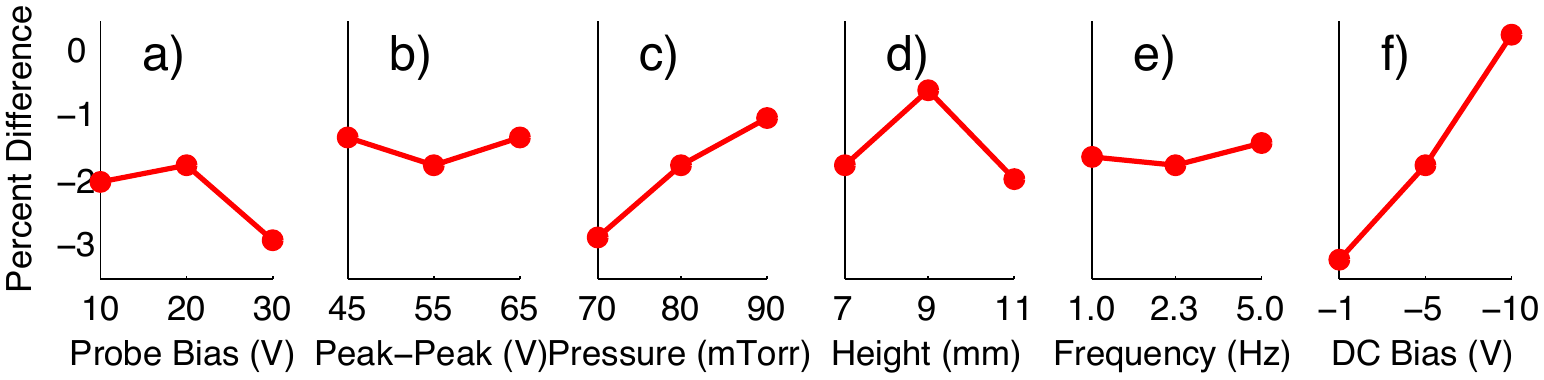}
\caption{(Color online) Percent difference of the predicted height from the experimental equilibrium position, based on the assumed charge for the base configuration.\label{oscamps2}}
\end{figure}

The initial force balance for the base configuration (as given in Table \ref{parameters}) is shown in Fig.~\ref{forcebalplot}, where the forces are normalized by the gravitational force. Application of the model under the parameter variation used is seen in Fig.~\ref{oscamps2}, where the predicted height (using Q=14,750e for each case) differs from the actual levitation height by less than 3\%. The charge on the particle was then adjusted so that the electric force balances the gravitational and ion forces at the experimental levitation height. This is 16,590e for the base configuration, which is 12\% larger than the initially assumed charge (generating a 2\% change in height). Note that the experimental charge must be found numerically because changing the grain charge changes the ion drag force, which is given by a transcendental function, which does not admit an analytical solution.

\begin{table}
\begin{tabular}{|c|ccc|c|ccc|}
\hline
Parameter & Low & Mid & High & Parameter & Low & Mid & High\\
\hline
Probe Bias & 14.9 & 16.6 & 18.2 & Height & 16.6 & 15.3 & 16.9\\
Peak to Peak & 16.1 & 16.6 & 16.1 & Frequency & 16.4 & 16.6 & 16.2\\
Power & 10.7 & 16.6 & 39.3 & DC Bias & 14.5 & 16.6 & 18.2\\
Pressure & 18.0 & 16.6 & 15.8 & --- & -- & -- & --\\
\hline
\end{tabular}
\caption{Dust particle charge found from the balance of the forces discussed in the text, in units of $10^3$ electron charges. All parameters were changed independently relative to the base configuration. The large particle charge for the highest power occurs due to the particle approaching the sheath edge.}
\label{chargetable}
\end{table}

Table \ref{chargetable} gives the resulting particle charges found for all parameters. As shown, the charge increases with power, DC bias, and probe bias, but decreases with pressure, consistent with the observed rise and fall in particle height, respectively. Calculating the percentage difference in charge as the remaining parameters are changed finds them to be consistent within 8\%.

The nonlinear increase in charge found with increasing power may be due to overestimation of the ion speed as mentioned previously. In the $P$=1.5 W case, the particle levitates near the sheath edge. A nonlinearity in the electric field \cite{douglass} occurs here which is not incorporated into the model. This results in an unrealistic increase in the ion drag. Note that the resulting charge increase with power is due to both the decrease in position of the sheath edge and the increase in the particle height as shown in Fig.~\ref{vertoscpos}. 

Comparing the experimental data with the model allows examination of the mechanism behind the asymmetric oscillation observed. There are several ways to perturb the electric field model in order to produce a vertical oscillation. First, the 1/$e$ point, when experimentally tracked over time, can be used to adjust the position of the sheath edge, which changes the electric field calculated by Eqn.~\ref{eforce}. Second, the potential at the sheath edge (i.e., the plasma potential) can be altered sinusoidally to follow the potential on the probe while the sheath edge remains fixed. Third, the charge of the particle can be changed by oscillating the plasma power, which also changes the ionization rate and plasma density. Since the change in plasma power and the manner in which it affects the grain charge requires many assumptions, the charge is instead estimated as a function of the grain's vertical position and the plasma emission intensity measured during probe potential oscillation. For each of these methods, the force balance can be used to predict the position of the particle over time. A change in the ion flow due to the probe potential was also considered. However, when the probe is more positive than the plasma potential, downward ion flow should increase in speed, reducing the particle levitation height. Therefore, it can not independently generate this oscillation.

\begin{figure}
\includegraphics[scale=0.626,trim=3 0 0 0]{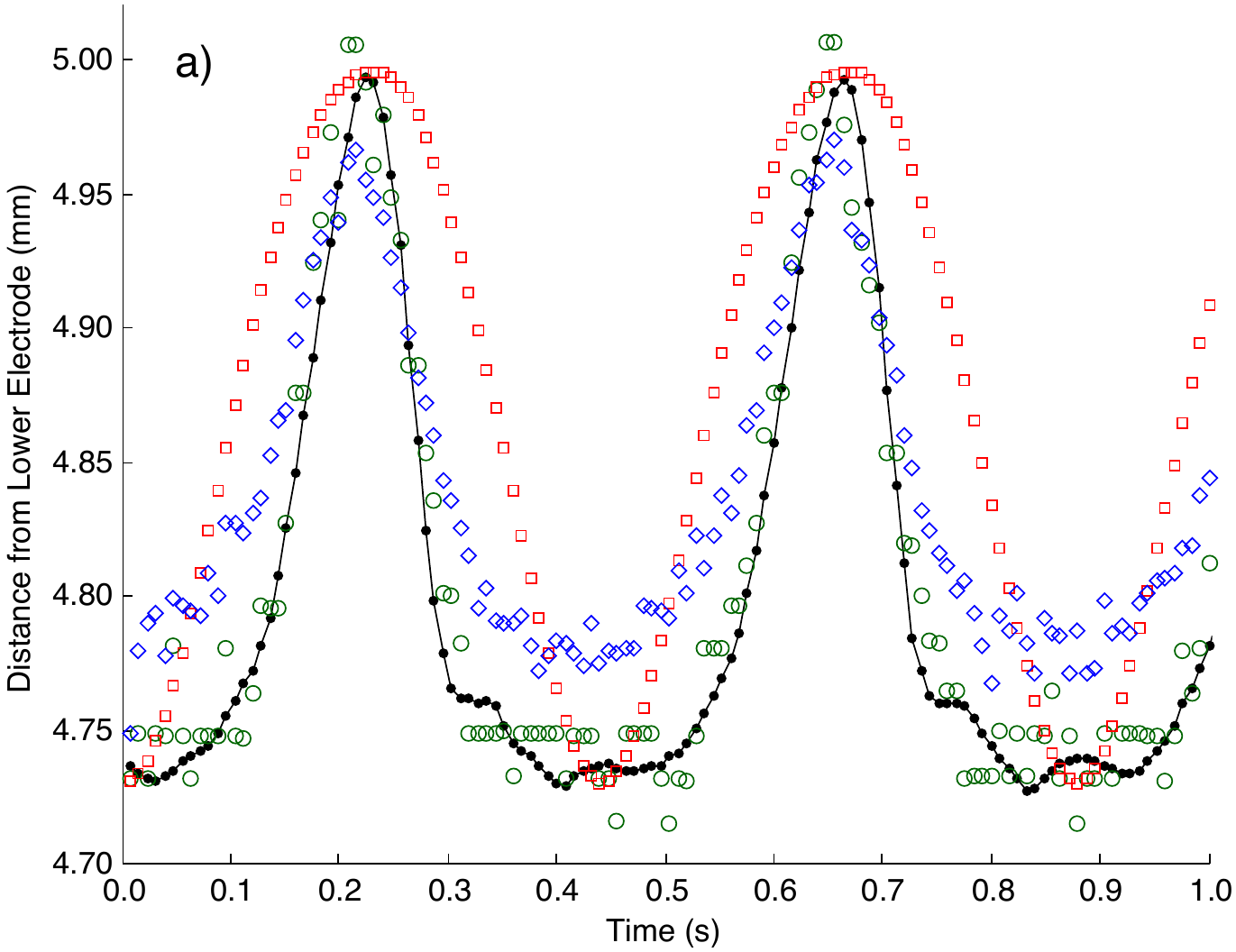}
\includegraphics[scale=0.641,trim=1 0 0 0]{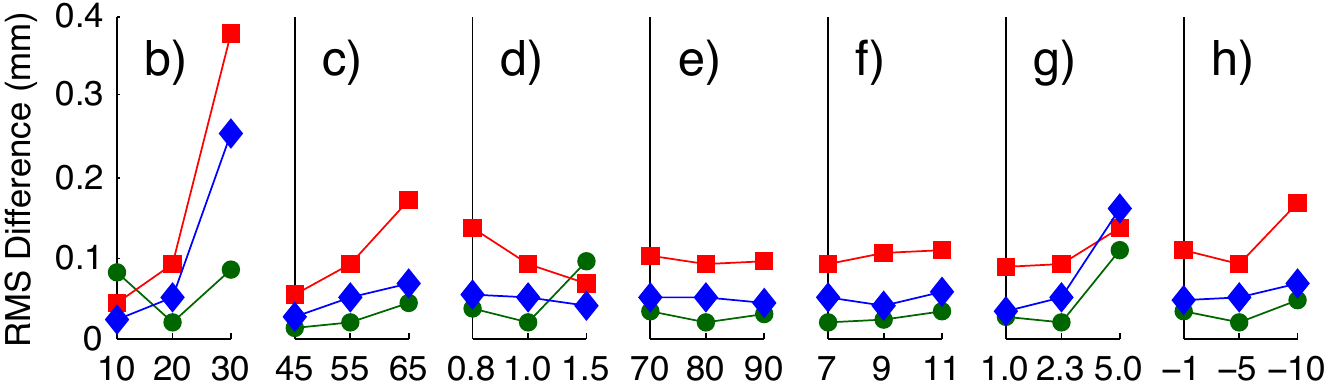}
\caption{(Color online) Predictions of multiple perturbations to the simple electric field model for the base configuration. a) The (black) dots give the the experimental particle position. The (red) squares represent the result of adjusting the sheath edge. The (blue) diamonds represent changing the plasma potential based on the probe oscillation. The (green) circles represent changing the grain charge (with the ion drag calculated from the equilibrium charge). The RMS deviation of the perturbations to the experimental data are shown in b)-h) where the parameter varied is b) probe bias (V), c) probe peak to peak (V), d) system power (W), e) pressure (mTorr), f) probe height (mm), g) frequency (Hz), and h) DC bias (V).\label{pertmeths}} 
\end{figure}

The results of the perturbations discussed above are shown in Fig.~\ref{pertmeths}, using the experimentally determined particle charge. As shown, calculations based on applying the experimental shift of the sheath edge yield the best prediction of the particle's motion (green circles). Taking the probe bias directly into account by applying a sinusoidal plasma potential results in sinusoidal particle motion (red squares). For the base configuration, the probe bias and amplitude used in the model had to be reduced from experimental values by about 50\% in order to match the measured equilibrium height and oscillation amplitude of the dust. This reduction is due to plasma shielding: a Debye length of 695 microns was calculated following the formula in \cite{popel} (which was used for the ion drag calculation). However, an increased shielding is expected since the probe is several millimeters away from the dust. Because the electron temperature change in the vertical direction was not estimated (such as in \cite{samarian}), shielding was not included. For the third perturbation, the motion is created by allowing the charge to change as a function of the maximum emission intensity, $I(t)$, resulting in motion indicated by the blue diamonds. In this case, the charge ranges from its equilibrium value ($Q_0$) to the value found from the force balance when the particle is at its maximum position ($Q_{max}$). This result is adjusted following an interpolation inversely proportional to the reduction in $I(t)$ shown in Fig.~\ref{intparticle}, whose functional form is
\begin{equation}
Q(I(t))=Q_0+(Q_{max}-Q_0)\frac{I_{max}-I(t)}{I_{max}-I_{min}},
\end{equation}
where $I_{min}$ ($I_{max}$) is the minimum (maximum) value over time of the maximum plasma intensity.
The ion drag calculation in this case is fixed, and uses only the equilibrium charge value.

Although a reduction in intensity generally signals a reduction in power delivered to the plasma and thus a reduced grain charge, the fact that the particle position increases implies the reverse. There are several possible explanations for this. Increasing the electron flow to the positive probe would increase the electron flux impacting the dust. The particle could also rise due to the increase in plasma shielding length due to a reduced electron density; this would increase direct attraction by the probe and repulsion from the lower electrode. The probe's electric field could interact with that of the sheath as well; however, since the tip is approximately 10 shielding lengths away from the dust, these effects are limited. Therefore, the most likely process is that the plasma bulk contracts due to a reduction in electrons as the probe collects them, raising the sheath edge, while the charge remains relatively constant.

The plots shown in Fig.~\ref{pertmeths} (b-h) illustrate the root mean square (RMS) deviations for the parameters tested as compared to experimental results. The RMS deviation for modifying the location of the sheath edge is smallest in 13 of the 15 cases, implying this is the best possible explanation. However, the other effects may contribute to the oscillation as well. Besides these perturbations, the speed of the flowing ions could be directly affected by the probe, but the ion drag would need to increase a few orders of magnitude in order to independently generate the observed particle amplitude, and the greatest motion would then occur during the maximum negative potentials (a phase shift of $\pi$ from that seen in the experiment). Therefore this oscillation can be considered a far field effect, unlike the near field effect studied in the powered horizontal wire experiment \cite{samsonov}.

\section{\label{sec:Con} Conclusions}

An adjustable, powered vertical probe was used to provide direct measurement of the neutral drag coefficient, the resonant frequency, and the charge on levitating particles in an rf powered dusty plasma. Measured values are comparable to those found in previously published experiments \cite{zhang, trottenberg}. The probe potential was oscillated to create vertical oscillations of the dust particles. The amplitude of the resulting dust oscillation was shown to be most affected by increasing the probe bias; this results in the probe potential spending more time above the plasma potential. Other parameters substantially altering the dust oscillation amplitude were determined to be the probe's peak to peak oscillation potential, the lower electrode DC bias, and the plasma power (Fig.~\ref{oscamps}).

A phase delay between the driving force and the maximum particle height was observed (Fig.~\ref{delay}). Such a phase delay is predicted for a forced, damped oscillator, with the calculation dependent on the forcing function. The measured particle oscillation amplitude was shown to follow that of a forced, damped harmonic oscillator, as seen by a fit to the amplitudes over a range of frequencies (Fig.~\ref{ampvsfreq}). Interestingly, the phase delay did not. Although a sine wave was applied to the probe, asymmetric oscillation of the particle and sheath edge showed the probe primarily interacts not directly with the particle but instead with the plasma.

The plasma discharge intensity was shown to decrease by a maximum of 9\% during probe oscillation (Fig.~\ref{intparticle}), signaling a reduction in the electron density in the plasma. The greatest decrease occurs in the bulk (Fig.~\ref{intcontour}). One unexpected result was that the position of the local maximum of the derivative of the emission profile in the lower sheath remained constant (Fig.~\ref{vertoscpos}) under variation in all system parameters.

The electric field in the sheath was studied under applied perturbation by analyzing the plasma glow. Under the low power regime examined, the 1/$e$ point of the emission profile was shown to be a useful measurement of the sheath edge, as corroborated by \cite{beckers} or confirmed when applied to another experiment \cite{samarian}. Calculating a force balance (Fig.~\ref{forcebalplot}) between the sheath electric field, gravity, and ion drag led to a predicted levitation height in agreement to that found in experiment (Fig.~\ref{oscamps2}). Thus, the greatest change in dust particle charge is caused by changes in plasma power, neutral gas pressure, and DC bias on the lower electrode.

Perturbations to the electric field model examined three possible ways of generating this oscillation: making the sheath height dynamic, changing the plasma potential, and adjusting the charge on the particle (Fig.~\ref{pertmeths}a). RMS calculations showed that changing the location of the sheath edge in time yielded the best fit to the collected particle data (Fig.~\ref{pertmeths}b-h).

Several areas of this experiment could be expanded. With appropriate calibration the intensity reduction observed in the bulk could be used to calculate the change in electron density. This could provide an extra parameter in plasma electromagnetic wave transmission studies, such as Faraday rotation. The fixed point of the position of the local maximum of the derivative of the emission profile may be purely dependent on the geometry of the cell and might be useful as a point of reference for image analysis. Theoretical work should be conducted to uncover what conditions can generate this result. Ion drag in the numerical model may be improved by changing the formulas used to calculate the decrease in ion and electron densities from the sheath edge \cite{land}, or testing the importance of the ion-neutral collisions and the nonlinearity near the sheath edge. Though the ion drag is small for most of the plasma parameters in this experiment, it does increase significantly with power and would increase if the ion speed was found to be greater than the minimum Bohm velocity.

\begin{acknowledgments}

The authors thank Jorge Carmona-Reyes, M.~Sc., Dr.~Angela Douglass, Dr.~Jie Kong, Jimmy Schmoke, and Mike Cook.

\end{acknowledgments}

\end{document}